# Impact of risk factors on work zone crashes using logistic models and Random Forest

**Huthaifa I Ashqar**
Booz Allen Hamilton
Washington, D.C., 20003 USA
Email: hiashqar@vt.edu

**Qadri H Shaheen**
School of Visual & Built Environments
Eastern Michigan University, Ypsilanti, MI, USA, 48197
Email: qshaheen@emich.edu

**Suleiman A Ashur**
School of Visual & Built Environments
Eastern Michigan University, Ypsilanti, MI, USA, 48197
Email: sashur@emich.edu

**Hesham A Rakha**
Department of Civil and Environmental Engineering
Virginia Tech, VA 24061 USA
Email: hrakha@vt.edu

**ABSTRACT**

Work zone safety is influenced by many risk factors. Consequently, a comprehensive knowledge of the risk factors identified from crash data analysis becomes critical in reducing risk levels and preventing severe crashes in work zones. This study focuses on the 2016 severe crashes that occurred in the State of Michigan (USA) in work zones along highway I-94. The study identified the risk factors from a wide range of crash variables characterizing environmental, driver, crash and road-related variables. The impact of these risk factors on crash severity was investigated using frequency analyses, logistic regression statistics, and a machine learning Random Forest (RF) algorithm. It is anticipated that the findings of this study will help traffic engineers and departments of transportation in developing work zone countermeasures to improve safety and reduce the crash risk. It was found that some of these factors could be overlooked when designing and devising work zone traffic control plans. Results indicate, for example, the need for appropriate traffic control mechanisms such as harmonizing the speed of vehicles before approaching work zones, the need to provide illumination at specific locations of the work zone, and the need to establish frequent public education programs, flyers, and ads targeting high-risk driver groups. Moreover, the Random Forest algorithm was found to be efficient, promising, and recommended in crash data analysis, specifically, when the data sample size is small.

**Keywords:** Work zone safety, risk factors, crash severity, machine learning, logistic regression.

# INTRODUCTION

Highway construction activities create work zones that are disruptive to regular traffic patterns and generate serious safety concerns [1]. Improving safety in work zones with minimum or no impact on the main function of highways has been a major goal of planners, traffic engineers and researchers. Increasing work zone safety is shaped by a large variety of risk factors of which some have not been studied well [2]. Traffic safety engineers must satisfy two contrasting functions, namely: protecting the workers at work zones to keep them safe and effectively maintaining a stable flow of traffic through the work zones all the time. Consequently, traffic safety engineers need to utilize historical work zone crash data to identify the risk factors that could lead to high-severity crashes injury and/or fatal crashes. Past studies identified several factors contributing to severe work zone crashes; however, crash characteristics are not universally the same for all highway sections. Hence, this study addresses the crash risk factors that could lead to high-severity crashes, injury and/or fatal crashes, along one of the major highways in the state of Michigan in the United States (US), namely, Interstate 94 (I-94). The I-94 is part of the US Interstate Highway System and starts at the City of Port Huron in the State of Michigan and runs west through the Great Lakes and terminates at the US northern Great Plains in the City of Billings in the State of Montana. The highway serves many areas including the Metro Detroit, connecting Michigan's largest city to its main airport in Romulus; 30 miles east of the City.

Every year the Michigan Department of Transportation (MDOT) exerts significant effort, time, and resources to maintain, rehabilitate, and/or construct new highways. This massive effort is expected to increase due to the deterioration of the roads as stated by the Michigan Section of the American Society of Civil Engineers [3]. The report gave a grade of D- for the roads and C- for bridges in Michigan and stated that *"Based on a 2016 assessment, 39% of Michigan's 120,000 miles of paved roadways are rated in poor condition, 43% are rated in fair condition, and just 18% are rated in good condition,"* and *"Approximately 1,234 bridges (11%) of the state's 11,156 bridges are structurally deficient."*

Generally, the Federal Highway Administration (FHWA) [4] reported that a work zone crash occurred every 5.4 minutes; every day, 70 work zone crashes resulted in at least one injury; and every week, 12 work zone crashes resulted in at least one fatality in 2015. About 96,626 crashes near work zones, an increase of 7.8% over 2014, constitute 2% of all roadway fatalities in the US in 2015. The severity of work zone crashes is classified into three major types, according to the damages and type of injuries resulting from these crashes. In 2015, records show that 73% of the crashes were property damage only, 26.4% resulted in at least one injury, and 0.7% resulted in at least one fatality. The fatality percentage for non-work zone crashes in 2015, were less than the work zone crashes by 0.2%.

In this study, we examined a wide range of crash variables along I-94 during construction in Michigan in 2016 characterizing environmental, driver, crash, and road-related variables. Through the examination, the risk factors that could lead to high-severity injury and/or fatal crashes were identified and their impact on crash severity was quantified. The impact of these risk factors on crash severity was investigated using frequency analyses, logistic regression statistics, and a machine learning algorithm, namely, Random Forest (RF). The results of this study will help traffic engineers to better understand the risk factors and their contribution to the likelihood of having fatalities when a severe crash occurs in a work zone. This will help planners and traffic engineers in improving Traffic Control Plans (TCPs) for work zones and use effective safety countermeasures and temporary traffic control devices to improve safety in work zones.

## RELATED WORK

In 2014, The National Highway Traffic Safety Administration (NHTSA) [5] reported a total of 607 fatal work zone crashes involving 669 fatalities in the State of Michigan with 23% of these crashes occurring in urban interstate work zones and 20% occurring in urban arterial work zones. Speeding was the major factor and caused 28% of the fatalities.

Identifying the significant risk factors near work zones has been studied previously using different approaches at different locations. Li and Bai [2] identified the significant risk factors based on a screening process that incorporated both statistical analyses and an empirical research approach. They systematically investigated risk factors using logistic regression and frequency analysis techniques. The researchers examined a wide range of crash variables characterizing work zone settings, environmental conditions, driver characteristics, and crash information using Kansas work zone crash data. Sze and Song [6] examined the effect of possible risk factors contributing to severe injury and fatality in work zone related crashes in New Zealand. A multinomial logistic regression model was developed to determine the association between crash severity and contributing factors such as the road environment, vehicle attributes, driver behavior, and crash circumstance. Moreover, Osman et al. [7] investigated the factors contributing to the severity of motorists' injuries in passenger-car crashes in different work zone configurations using a Mixed Generalized Ordered Response Probit (MGORP) modeling framework. Meng et al. [8] also developed a novel probabilistic Quantitative Risk Assessment (QRA) model to evaluate the casualty risk combining frequency and consequences of all crash scenarios triggered by long-term work zone crashes. Furthermore, Ghasemzadeh and Ahmed [9] used a probit–classification tree to identify factors affecting work zone crash severity in adverse weather conditions using eight years of work zone weather-related crashes (i.e., 2006–2013) in the State of Washington. Likewise, Koilada [10] examined and identified factors that influence crash injury severity in work zone areas using five years (2010-2014) of crash data from North Carolina.

This study has four main goals. First, update the information of 2016 crash data and worker injuries and fatalities near work zones on I-94 in Michigan. Second, identify risk factors contributing to severe crashes at work zones. Third, analyze the risk factors using a machine learning algorithm, which has not been used before to analyze crash severity, and compare the results with the traditional statistical analysis findings. Finally, develop recommendations to improve safety at work zones for motorists and workers before and during the construction project time.

## DATA SET

A major step in this research is to collect the crash data and related factors in work zones along I-94 in Michigan. This step was achieved by utilizing MDOT maps and Michigan State Police crash data databases. The MDOT Geographic Information System (GIS) was used to locate and analyze the work zones by identifying the location and duration of 2016 construction projects on the Michigan highway system. On the other hand, the state police department provides the nearest coordinates of the crash location for every reported crash. Using the Michigan Traffic Crash Facts website [11], a built environment query was built based on two major factors: period and geographic location based on the road segment. The captured data could be filtered based on the required information. By locating these crashes using GIS on the MI highway map, and by buffering the digitized path of the construction zone, information about a crash that occurred on a determined area in certain times was extracted. More information can be found in [12].

This study focuses on the 2016 severe crashes at four construction locations, as shown in Figure 1. We identified the at-fault drivers in the original crash database and compiled their characteristics and other necessary crash information into spreadsheets utilizing the state crash database. The collected crashes and crash-related information were organized including various variables with specific observations. Table 1 lists the highway maintenance work type, construction zone length, period, and number of crashes during construction for the four different work zones on highway I-94. It is worth nothing that although location 2 is the longest one, there were no crashes occurred near it during the construction time.

Table 1 Summary of Four Selected Construction Locations.

| ID[a] | Work Type | Length (mi) | Start | End | # of Crashes |
|---|---|---|---|---|---|
| 1 | Resurface | 7.4 (11.9 km) | Mar. | Oct. | 67 |
| 2 | Resurface | 8.1 (13.0 km) | Sept. | Oct. | n/a[b] |
| 3 | Patch Pavement | 2.4 (3.90 km) | May. | Aug. | 76 |
| 4 | Resurface | 5.6 (9.01 km) | Mar. | Nov. | 236 |

[a] Shown in Figure 1.
[b] No crashes during construction.

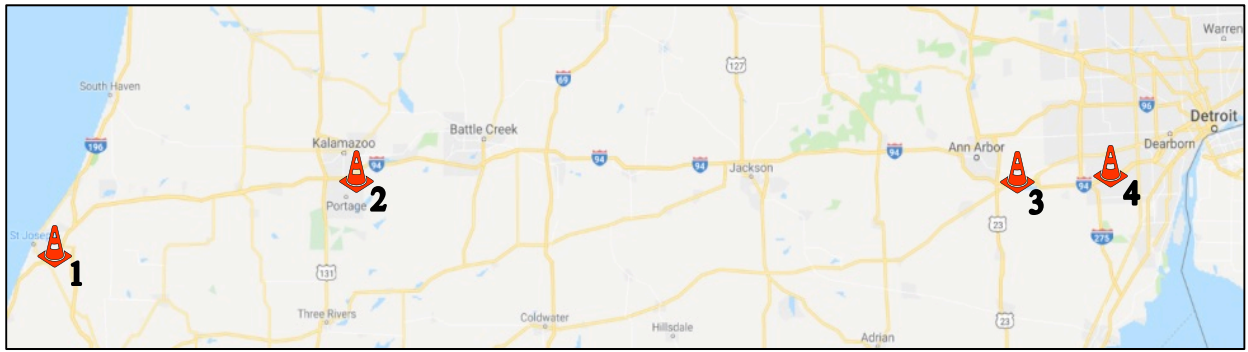

Figure 1 Location of the four selected locations at I-94 in Michigan [13].

## METHODS

### Crash Frequency and Severity Index

Crash frequency is defined as the total number of crashes occurring in a location at a determined work zones setting. This study used the Severity Index as a measure of fatal, injury and property damage crashes [15, 16]. A crash severity index gives an indication of the crash severity at a location and is computed using Equation 1 [14]. As shown in the equation, fatal crashes are weighted 4 times injury crashes and 12 times Property-Damage Only (PDO) crashes.

$$\textbf{\textit{Severity Index}} = \frac{12 \times F + 3 \times I + PDO}{N} \qquad (1)$$

where;
$F$ = the total number of fatal crashes at a particular location;
$I$ = the total number of injury crashes at a particular location;
$PDO$ = the total number of property-damage crashes at a particular location; and
$N$ = the total number of crashes at a particular location.

**Logistic Regression**

This study used logistic regression to assess the impact of work zone risk factors. Logistic regression is used to describe data and to explain the relationship between one dependent variable (Severity Index) and a wide range of work zone crash variables at the four construction sites along I-94. The significance of this technique in traffic safety-related studies has been recognized for years [2, 17-19].

**Random Forest (RF)**

Breiman proposed RF as a new classification and regression technique in supervised learning [20]. RF creates an ensemble of decision trees and randomly selects a subset of features to grow each tree. While the tree is being grown, the data are divided by employing a criterion in several steps or nodes. The correlation between any two trees and the strength of each individual tree in the forest affect, also known as the forest error rate in classifying each tree. Practically, the mean squared error of the responses is used for regression.

RF offers several advantages [20, 21]. For example, there are very few assumptions attached to its theory; it is considered to be robust against overfitting; it runs efficiently and relatively quickly with a large or small amount of data and many input variables without the need to create extra dummy variables; it can handle highly nonlinear variables and categorical interactions; and it ranks each variable's individual contribution to the model. However, RF also has a few limitations. For instance, the observations must be independent, which is assumed in our case.

RF was used in this study by applying the 'randomForest' package in R [22] to rank the different risk factors based on their relative importance (i.e. percent increase in Mean Square Error (%IncMSE)). %IncMSE is computed from permuting out-of-bag (OOB) data in which for each tree, the prediction error on the OOB portion of the data is recorded (i.e. error rate for classification, or MSE for regression). Then the same is done after permuting each predictor variable. The difference between the two are then averaged over all trees and normalized by the standard deviation of the differences. Although using the %IncMSE to rank risk factors might not reveal the significance of a specific factor as logistic regression, it exposes a better understanding of the risk factors and how they could increase the likelihood of having fatalities when a severe crash occurs in a work zone. It might be also a suitable use for prediction purposes, which is beyond the scope of this study.

Crash data suffer from some problems such as being characterized by a small number of observations, under-reporting, and non-linear relationship bias between crash indices and explanatory variables. These issues are a potential source of error in modeling crash data that may cause incorrect estimates and inferences [23]. It is therefore crucial to implement machine learning to identify potential explanatory variables and compare them with the results from logistic analysis. Given the aforementioned advantages of RF, it was used to rank the importance of the environment and driver categories at the variable's level. We chose 10,000 trees to grow in the RF, to ensure that every variable gets predicted sufficient times.

**ANALYSIS AND RESULTS**

The identified work zone risk factors were studied using crash frequency, logistic regression, and RF to understand their impact on the crash severity at four construction sites along I-94 in Michigan. The effect of work zone crash factors was assessed using logistic regression and RF to expose their impact on causing fatalities, injuries, and property damages when a severe crash occurred. These results will benefit work zone traffic control designers and managers in providing real world insights for reducing high-risk factors in work zones and consequently mitigating crash severity and frequency.

**Environment Information**

Statistical test showed that all the environment variables were significant risk factors affecting crash severity as shown in Table 2. Moreover, using machine learning algorithm, RF, showed that roadway lighting conditions are the most important variables followed by weather conditions. Although more than 68% of the crashes occurred in a good light condition, about 14% of them occurred in a fair condition and about 11% occurred in poor conditions. These results indicate that poor light conditions could increase the probability of causing a severe crash that result in fatalities and/or injuries. In this analysis, "good light condition" refers to daylight, "fair condition" refers to the dawn, dusk, or dark-with-streetlights, and "poor condition" refers to the dark-without-streetlights. In terms of weather conditions, rain was not a significant factor affecting crash severity, while clear, cloudy, and snow conditions were significant risk factors. This indicates that unlike rain, snow could increase the probability of causing fatalities and/or injuries when a severe crash occurs. Lighting conditions and weather conditions at the construction site were reported for most of the crashes. However, information about whether a deer was involved/associated in a crash was not reported for about 98% of the crashes, although it seems to have a significant impact on the crash severity based on a small percent of information reported (less than 2%).

**Driver Information**

This category mainly describes the at-fault driver age and driver error associated with the crash such as texting, drinking alcohol (DUI), and drug use. Although the statistical (logistic regression) analysis could not establish significant models for the probability of having fatalities and/or injuries with at-fault driver age, crash frequencies showed that young drivers (<24 years old) were involved in more crashes than senior drivers (>60 years old). Nevertheless, about 50% of the crashes were contributed by drivers between the ages of 24 and 60. Statistical tests also show that driver errors are significant risk factors affecting crash severity, as shown in Table 2.**Error! Reference source not found.**Whether the at-fault driver was distracted, DUI, or using drugs are all risk factors that have significant impact on crash severity. In terms of the RF results, the most important factor that could increase the crash severity is when the driver is distracted followed by DUI. Although Drug use is found to be significant using logistic regression, it is the least important factor using the RF analysis.

Table 2 Crash Frequency, Significance, and Importance of Crash Information Variables.

| Category | Variable | Observation | Crash Frequency | Statistical significance | | Importance | |
|---|---|---|---|---|---|---|---|
| | | | | *p-value* | *Significant?* | *%IncMSE* | *Rank* |
| **Crash Information** | ***Total Units Reported*** | 1 | 32.98% | 0 | Yes | 56.10 | 3rd |
| | | 2 | 62.53% | 0.001 | Yes | | |
| | | 3 | 2.90% | 0.001 | Yes | | |
| | | 4 | 1.58% | 0 | Yes | | |
| | ***Relation to Roadway*** | In the gore | 0.26% | 0.99 | No | 53.34 | 4th |
| | | In the median | 4.75% | 0.99 | No | | |
| | | On the road | 79.16% | 0.99 | No | | |
| | | On the shoulder | 8.71% | 0.99 | No | | |
| | | Other / unknown relationship | 2.11% | 0.99 | No | | |
| | | Outside of the shoulder/curb-line | 5.01% | 0.99 | No | | |
| | ***Crash Type*** | Angle | 1.32% | 0.66 | No | 88.03 | 1st |
| | | Head-on / left turn | 0.26% | 0.99 | No | | |
| | | Other | 3.96% | 0.99 | No | | |
| | | Rear-end | 37.20% | 0.089 | Yes | | |
| | | Rear-end left turn | 0.26% | 0.99 | No | | |
| | | Sideswipe opposite direction | 0.26% | 0.99 | No | | |
| | | Sideswipe same direction | 25.33% | 0.015 | Yes | | |
| | | Single motor vehicle | 31.40% | 0.17 | No | | |
| | ***Motorcycle*** | Motorcycle involved | 0.53% | 0 | Yes | 59.52 | 2nd |
| | | No motorcycle involved | 99.47% | 0 | Yes | | |
| | ***Truck or Bus*** | No truck or bus involved | 84.17% | 0 | Yes | 33.88 | 5th |
| | | Truck or bus involved | 15.83% | 0.19 | No | | |
| | ***Lane Departure*** | Multiple vehicle | 0.26% | 0 | Yes | 29.05 | 6th |
| | | No lane departure | 73.09% | 0.63 | No | | |
| | | Parked vehicle | 0.26% | 0.044 | Yes | | |
| | | Single vehicle | 26.39% | 0.42 | No | | |
| | ***Investigated at Scene*** | Investigated at scene | 88.65% | n/a | n/a | n/a | n/a |
| | | Not investigated at scene | 11.35% | n/a | n/a | | |
| | ***Crash: Hit-and-run*** | Hit-and-run | 11.08% | n/a | n/a | n/a | n/a |
| | | Not hit-and-run | 88.92% | n/a | n/a | | |
| | ***Crash: Fleeing/Pursuing*** | No pursuit | 99.74% | n/a | n/a | n/a | n/a |
| | | Pursuit involved | 0.26% | n/a | n/a | | |
| | ***Fatal Crash*** | No fatality (may be injuries) | 99.47% | n/a | n/a | n/a | n/a |
| | | Fatal crash (may also be injuries) | 0.53% | n/a | n/a | | |
| | ***Worst Injury in Crash*** | Fatal injury (K) | 0.53% | n/a | n/a | n/a | n/a |
| | | No injury (O) | 76.78% | n/a | n/a | | |
| | | Possible injury (C) | 13.19% | n/a | n/a | | |
| | | Suspected minor injury (B) | 6.86% | n/a | n/a | | |
| | | Suspected serious injury (A) | 2.64% | n/a | n/a | | |
| | ***Injury crash*** | Injury crash (no fatalities) | 22.69% | n/a | n/a | n/a | n/a |
| | | No injuries (may be fatality) | 77.31% | n/a | n/a | | |
| | ***Property Damage*** | No property damage | 23.22% | n/a | n/a | n/a | n/a |
| | | Property damage involved | 76.78% | n/a | n/a | | |
| **Environment Information** | ***Lighting conditions*** | Dark lighted | 14.51% | 0 | Yes | 101.29 | 1st |
| | | Dark unlighted | 11.35% | 0 | Yes | | |
| | | Dawn | 4.22% | 0.009 | Yes | | |
| | | Daylight | 68.07% | 0.0001 | Yes | | |
| | | Dusk | 1.32% | 0.3 | No | | |
| | | Unknown | 0.53% | n/a | n/a | | |
| | ***Weather conditions*** | Clear | 58.58% | 0 | Yes | 68.96 | 2nd |
| | | Cloudy | 22.16% | 0.088 | Yes | | |
| | | Fog | 0.53% | 0.39 | No | | |
| | | Rain | 10.03% | 0.57 | No | | |
| | | Sleet/hail | 0.53% | 0.55 | No | | |
| | | Snow | 7.39% | 0.001 | Yes | | |
| | | Unknown | 0.79% | n/a | n/a | | |
| | ***Deer associated*** | Deer involved | 1.85% | 0 | Yes | 45.53 | 3rd |
| | | Uncoded errors | 98.15% | n/a | n/a | | |
| **Driver Information** | ***Driver age*** | 16 | 0.53% | 0.43 | No | 25.72 | 3rd |
| | | 17 | 2.11% | 0.44 | No | | |
| | | 18-20 | 10.03% | 0.62 | No | | |

| Category | Variable | Observation | Crash Frequency | Statistical significance | | Importance | |
|---|---|---|---|---|---|---|---|
| | | | | *p-value* | *Significant?* | *%IncMSE* | *Rank* |
| | | 21-24 | 19.79% | 0.59 | No | | |
| | | 24-60 | 50.40% | 0.57 | No | | |
| | | 60-64 | 5.80% | 0.88 | No | | |
| | | 65-74 | 9.76% | 0.81 | No | | |
| | | 75-84 | 1.58% | 0.78 | No | | |
| | ***Driver distracted*** | No | 94.99% | 0.019 | Yes | 46.42 | 1st |
| | | Yes | 5.01% | 0 | Yes | | |
| | ***DUI*** | Yes | 3.17% | 0 | Yes | 38.10 | 2nd |
| | | No | 96.83% | 0.051 | Yes | | |
| | ***Drug use*** | Yes | 1.32% | 0 | Yes | 9.79 | 4th |
| | | No | 98.68% | 0.26 | No | | |
| **Road Information** | ***Number of Traffic Lanes*** | 1 | 0.79% | 0.54 | No | 70.80 | 2nd |
| | | 2 | 13.98% | 0.65 | No | | |
| | | 3 | 70.98% | 0.45 | No | | |
| | | 4 | 10.03% | 0.23 | No | | |
| | | 5 | 0.79% | 0.53 | No | | |
| | | 6 | 3.43% | 0.22 | No | | |
| | ***Area of Road at Crash*** | Freeway Crash - All other freeway areas | 82.06% | 0 | Yes | 24.71 | 6th |
| | | Freeway Crash - Entrance/exit ramp related | 2.90% | 0.19 | No | | |
| | | Freeway Crash - Transition area (increase or decrease in travel lanes) | 3.43% | 0.33 | No | | |
| | | Non-Freeway Intersection Crash (Related to or within 150' of an intersection) - Within intersection | 0.53% | 0.99 | No | | |
| | | Other Non-Freeway Areas - Curved roadway not related to other selections | 0.53% | 0.32 | No | | |
| | | Other Non-Freeway Areas - Median crossing related | 0.26% | 0.99 | No | | |
| | | Other Non-Freeway Areas - Straight roadway not related to other selections | 9.50% | 0.025 | Yes | | |
| | | Other Freeway Crash - Curved roadway | 0.79% | 0.99 | No | | |
| | ***Road Conditions*** | Dry | 72.82% | 0 | Yes | 76.83 | 1st |
| | | Ice | 4.49% | 0.0025 | Yes | | |
| | | Slush | 0.53% | 0.88 | No | | |
| | | Snow | 4.22% | 0.0058 | Yes | | |
| | | Wet | 17.41% | 0.095 | Yes | | |
| | | other | 0.53% | n/a | n/a | | |
| | ***Speed Limit at Crash Site*** | 30 | 0.26% | 0.99 | No | 12.60 | 7th |
| | | 35 | 0.26% | 1 | No | | |
| | | 40 | 0.26% | 0.99 | No | | |
| | | 45 | 3.17% | 0.99 | No | | |
| | | 55 | 0.79% | 0.99 | No | | |
| | | 60 | 11.35% | 0.99 | No | | |
| | | 70 | 82.85% | 0.99 | No | | |
| | | Uncoded & errors | 1.06% | n/a | n/a | | |
| | ***Traffic Control*** | Signal | 0.53% | 0.014 | Yes | 0.01 | 9th |
| | | Stop sign | 0.53% | 0.7 | No | | |
| | | Yield sign | 0.53% | 0.008 | Yes | | |
| | | None | 98.42% | 0 | Yes | | |
| | ***Traffic Way*** | Divided highway, median strip, with traffic barrier | 73.88% | 0 | Yes | 46.55 | 4th |
| | | Divided highway, median strip, without traffic barrier | 20.58% | 0.22 | No | | |
| | | Nontraffic | 0.26% | 0.99 | No | | |
| | | Not physically divided (two-way trafficway) | 3.69% | 0.88 | No | | |
| | | One-way trafficway | 1.58% | 0.031 | Yes | | |
| | ***Intersection*** | Not intersection crash | 99.47% | 0.016 | Yes | 1.21 | 8th |
| | | Intersection crash | 0.53% | 0 | Yes | | |
| | ***Construction Crash Location*** | Before first warning sign | 0.53% | 0 | Yes | 45.69 | 5th |
| | | Between first and last sign | 14.78% | 0.92 | No | | |
| | | No warning signs | 0.53% | 0.022 | Yes | | |
| | | Uncoded & errors | 84.17% | n/a | n/a | | |
| | | Intermittent / moving work | 0.26% | 0.99 | No | 61.09 | 3rd |

| Category | Variable | Observation | Crash Frequency | Statistical significance | | Importance | |
|---|---|---|---|---|---|---|---|
| | | | | *p-value* | *Significant?* | *%IncMSE* | *Rank* |
| | *Construction Activity* | Lane closure | 12.14% | 0.99 | No | | |
| | | Lane shift / crossover | 2.11% | 0.99 | No | | |
| | | Other | 0.53% | 0.99 | No | | |
| | | Uncoded & errors | 81.53% | 0.99 | No | | |
| | | Work on shoulder / median | 3.43% | 0.99 | No | | |

**Crash Information**

Crash information contains 20 different variables. The significance and importance of the total units reported, relation to roadway, crash type, lane departure, and if the crash involved a motorcycle, a truck, and a bus are quantified. In contrast, the significance and importance of other recorded variables such as how many fatalities, injuries, PDO and the worst injury reported in a crash were not quantified. Given that these variables were used to calculate the crash severity index; they could produce biases if considered as independent variables. Moreover, other variables namely, crash investigated at scene, crash: hit-and-run, and crash: fleeing/pursuing were not also quantified because it does not make sense to investigate their effect as they are not related to the severity of a crash. Nonetheless, results show that there are about 11.35% of crashes not investigated at the scene, 11.08% of crashes of hit-and-run type, and relatively low percent (0.26%) of pursuit involved crashes. Investigating, collecting, and using accurate and comprehensive data related to road crashes are necessary not only for analysis but also for in-depth studying to identify the contributory factors to have a better understanding of the "chain-of-events" [24]. Thus, a detailed and thorough investigation, and information required for scene reconstruction are recommended.

Table 2 shows the crash frequency, the significance, and the importance of remaining variables. In general, the total units reported, the type of vehicles involved in the crash, and the crash type are highly correlated with crash severity. Rear-end and sideswipe same direction types of crashes were found to be significant. This inherently means at least two vehicles were involved in the crash, which corresponds to the highest crash frequency in the total units reported; two units. Statistical tests also show that crashes are more likely to be severe if no truck, no bus, and no motorcycle were involved in the crash. The term "truck" refers to such heavy vehicle types as single-unit large trucks, truck and trailers, tractor-trailers. Although parked vehicle-involved crashes and multiple vehicle crashes were not frequent in the data, they were found to be significant.

In terms of RF results, crash type was found to be the most important variable followed by if a motorcycle was involved in the crash, whereas lane departure was found to be the least important variable. Although six crash types out of eight was found to be statistically insignificant, RF quantified it as the most important variable. We hypothesize that statistical tests occasionally fail in dealing with small sample sizes, as is the case with the data used in this study, because the maximum likelihood estimation of parameters could produce insufficient results [25]. Nonetheless, RF has not only a unique advantage in dealing with small sample sizes, but also a potential for providing greater diversity within the forest [26, 27].

**Road Information**

Road information variables such as number of lanes, speed limit, road condition, area of read at crash, traffic control, and traffic way is important to be studied as it might have an impact on the crash severity. Table 2 shows the crash frequency and the significance of each variable. While it shows that the number of traffic lanes, speed limit, and the construction activity are not significant to mitigate crash severity, other variables were found to be significant and some of them should be studied in detail. One significant observation is that when the crash area is a straight roadway and not a ramp or a median crossing, it has an impact on the severity of crashes. Road condition and traffic control observations seem to be the most significant factors on the crash severity and the most important factor resulted from the RF analysis. When the road is wet or covered by snow, water creates a barrier between the road and tires cause driver to lose traction and/or glide. This is related to the traffic control at the work zone, in which results show a 98.42% of crash frequency when there is no traffic signal, and it has an impact on the crash severity. In terms of the traffic way observations, when the highway is divided, median strip, with traffic barrier and/or a one-way traffic lane, those are the two conditions that have a significant impact on crash severity. This is linked to the crash type analysis we discussed in the previous section, which rear-end type of crashes has a very high frequency nearby work zone locations. Although about 84% of the information on the crash location relative to the location of the warning signs were un-coded and/or errors, it is worth mentioning that the higher frequency (about 14%) of the crashes were reported between the first and last warning sign, which means it is directly related to the construction environment not to the traffic spilled back for example or other indirect consequences of the construction. It is also appealing to note that most of the information in some of these categories is either uncoded or has some errors when producing the data. These variables are very important and are recommended to be reported in the future.

**CONCLUSIONS**

The need to fix, maintain and rehabilitate highway infrastructure has been increasing in the US and is expected to increase due to the deterioration of these facilities over the years. The main challenge to the US Departments of Transportation and other regional, state, and local transportation agencies is to run the construction operations effectively while minimizing the impact of the construction of work zone operations and safety for travelers and construction workers. Work zone safety is affected by many risk factors and some of them might have not been fully understood by traffic engineers or might not be the same for different highways. Comprehensive knowledge of the risk factors discovered from crash data therefore becomes critical in reducing risk levels and preventing severe crashes in work zones. This study focuses on the severe crashes that occurred in work zones along I-94 in the state of Michigan, USA in 2016. This study identified the risk factors from a wide range of crash variables using logistic regression and RF. The findings of this study are valuable to traffic engineers for developing countermeasures in work zones that can alleviate the safety risk resulting from a wide range of factors among which some could be overlooked when designing and setting work zones.

Results show that when the crash type is rear-end or sideswipe in the same direction, it has higher probability to be severe. As a result, it is recommended that traffic engineers favorably weigh the need for traffic control mechanisms to harmonize the speed of vehicles before approaching work zones. It is important to note that a significant number of crashes were not investigated at the scene. These variables are very important and are recommended to be reported in the future to provide insightful recommendations, exploration, and reconstruction. Furthermore, involvement of a motorcycle in a crash was found the second most important factor in characterizing a crash as severe, thus, the researchers recommend a need for frequent public education programs, flyers, and ads orienting motorcyclists or other drivers in areas where motorcyclists are frequent.

In terms of environmental factors, results show that poor light conditions on the road could increase the probability of fatalities and/or injuries when a severe crash occurs. Thus, it is recommended that traffic safety engineers favorably weigh the needs for illumination giving drivers enough time to be aware of a construction zone. With regards to driver-related factors, the study demonstrates that at-fault drivers distracted or driving under the influence are the most important risk factors and have significant impact on crash severity. Consequently, the researchers recommend a need for frequent public education programs, flyers, and ads orienting these high-risk driver groups and a need to develop traffic enforcement strategies that result in better compliance rates.

It is important to notice that, in some cases, more than 80% of the reported crashes does not have some important information (e.g. the construction activities, and the crash location relative to the location of the warning signs). These variables are very important and are recommended to be reported in the future to conclude insightful recommendations and exploration. That said, it is worth mentioning that the higher frequency (about 14%) of the crashes were reported between the first and last warning sign, which means it is directly related to the construction environment not to the traffic spilled back for example or other indirect consequences of the construction.

We also used RF as a machine learning regression algorithm, beside the traditional-used logistic regression. We compared the results of RF and logistic regression in identifying potential explanatory variables. RF is found to be useful, promising, and recommended in crash data analysis, as they overcome the limitations of traditional statistical methods. Specifically, RF was useful in producing meaningful results in cases of small sample sizes in contrast to logistic regression. Moreover, we also recommend using RF as a predictive model for crash data.

## ACKNOWLEDGMENTS

The authors thank the Michigan Department of Transportation and the State Police Department for providing the needed information for this research.

## AUTHOR CONTRIBUTION STATEMENT

The authors confirm contribution to the paper as follows: study conception and design: H. Ashqar, Q. Shaheen, S Ashur, and H. Rakha; data collection: Q. Shaheen; analysis and interpretation of results: H. Ashqar; draft manuscript preparation: H. Ashqar, Q. Shaheen, S Ashur, and H. Rakha. All authors reviewed the results and approved the final version of the manuscript.